\documentclass{entcs} 
\usepackage{amsmath} 
\usepackage{amssymb}
\usepackage{graphicx}
\usepackage[colorlinks=true, linkcolor=red, citecolor=blue,urlcolor=green]{hyperref}

\def\anp#1#2#3{Annals Phys. #1 (#3) #2}
\def\arnps#1#2#3{Ann.\ Rev.\ Nucl.\ Part.\ Sci.\ #1 (#3) #2}

\def\cmp#1#2#3{Comm. Math. Phys. #1 (#3) #2}

\def\epja#1#2#3{Eur. Phys. J. A #1 (#3) #2}
\def\epjc#1#2#3{Eur. Phys. J. C #1 (#3) #2}
\def\ibid#1#2#3{{\it ibid.} #1 (#3) #2}
\def\ijma#1#2#3{Intl. Jour. Mod. Phys. A #1 (#3) #2}
\def\ijmpe#1#2#3{Intl. Jour. Mod. Phys. E #1 (#3) #2}
\def\ijtp#1#2#3{Intl. Jour. Theor. Phys. A #1 (#3) #2}
\def\jhep#1#2#3{J. High Energy Phys. #2 (#3) #1}

\def\jpcs#1#2#3{J.\ Phys.\ Conf.\ Ser.\ #1 (#3) #2}
\def\jpg#1#2#3{Jour. Phys. G #1 (#3) #2}
\def\mpla#1#2#3{Mod. Phys. Lett. A #1 (#3) #2}
\def\nat#1#2#3{Nature #1 (#3) #2}
\def\npa#1#2#3{Nucl. Phys. A #1 (#3) #2}
\def\npb#1#2#3{Nucl. Phys. B #1 (#3) #2}

\def\plb#1#2#3{Phys. Lett. B #1 (#3) #2}
\def\prc#1#2#3{Phys. Rev. C #1 (#3) #2}
\def\prd#1#2#3{Phys. Rev. D #1 (#3) #2}
\def\prl#1#2#3{Phys. Rev. Lett. #1 (#3) #2}
\def\phr#1#2#3{Phys. Rep. #1 (#3) #2}

\def\ptp#1#2#3{Prog. Theor. Phys. #1 (#3) #2}
\def\ptps#1#2#3{Prog. Theor. Phys. Supp. #1 (#3) #2}

\def\rmp#1#2#3{Rev. Mod. Phys. #1 (#3) #2}

\begin{document}

\begin{frontmatter}

\title{Phases of Dense Quarks at Large $N_c$}

\author[BNL,RBRC]{Larry McLerran} and \author[BNL]{Robert D. Pisarski}

\address[BNL]{Physics Department, Brookhaven National Laboratory, Upton, NY 11973, USA} 
\address[RBRC]{RIKEN BNL Research Center, Brookhaven National Laboratory, Upton, NY 11973, USA}

\begin{abstract}
In the limit of a large number of colors, $N_c$, we suggest that
gauge theories can exhibit several distinct phases
at nonzero temperature and quark density.  Two are familiar:
a cold, dilute phase of confined hadrons,
where the pressure is $\sim 1$, and a 
hot phase of deconfined quarks and gluons, with
pressure $\sim N_c^2$. 
When the quark chemical potential $\mu \sim 1$,
the deconfining transition temperature, $T_d$, is independent of $\mu$.
For $T < T_d$, as $\mu$ increases above the mass threshold,
baryons quickly form a dense phase where the pressure is $\sim N_c$.
As illustrated by a Skyrme crystal, chiral symmetry can be both spontaneously
broken, and then restored, in the dense phase.
While the pressure is $\sim N_c$, like that of (non-ideal) quarks, 
the dense phase is still confined,
with interactions near the Fermi surface 
those of baryons, and not of quarks.
Thus in the chirally symmetric region, baryons near the
Fermi surface are parity doubled.
We suggest possible implications for the phase diagram of QCD.
\end{abstract}
\end{frontmatter}

\section{Introduction}

Many of the observed properties of QCD can be understood, at least
qualitatively, by generalizing from three to a large 
number of colors, $N_c \rightarrow \infty$.
\cite{largeN,witten,baryons}.
For example, consider the phase transition at a nonzero temperature, $T$
\cite{thorn}.
At low temperature confinement implies that all states are color singlets, 
such as mesons and glueballs, with a pressure
$\sim N_c^0 \sim 1$.  At high temperature gluons 
in the adjoint representation deconfine, contributing
$\sim N_c^2$ to the pressure.
Thus one can define the deconfining transition
simply by the point where the term $\sim N_c^2$ in the pressure turns
on.  The transition temperature for deconfinement,
$T_d$, is expected to be of order one at large
$N_c$, on the order of a typical QCD mass scale, such as the renormalization
mass parameter, 
$\Lambda_{QCD} \sim 200$~MeV.  Arguments suggest that deconfinement
is a strongly first order transition, with a latent
heat $\sim N_c^2$.  Since the free energy of
$N_f$ flavors of deconfined quarks is $\sim N_c N_f$
in the limit of large $N_c$, and small $N_f$, 
deconfinement probably
drives chiral symmetry restoration at $T_d$.
Several of these features have been confirmed by
numerical simulations on the lattice \cite{teper}.

In this paper we consider the phase diagram in the plane of temperature
and quark chemical potential, 
$\mu$ 
\cite{dense,two_trans,color_super,cs_pert,cs_crystal,njl,chir_dens,tricritical,cleymans,random,bielefeld96,lattice,lattice_mu,two_colors,schon_thies,banks_ukawa,bielefeld_ren_loop,sch_dyson,pert_temp}.
It is usually assumed that for all $T$ and $\mu$,
there is a single transition, at which
both deconfinement and chiral symmetry restoration occurs,
shaped something like a semi-circle.
If the transition is crossover for $\mu = 0$ and $T \neq 0$ \cite{lattice},
there could be a chiral critical end point in the plane of $T$ and $\mu$
\cite{tricritical}.  
From numerical simulations on the lattice \cite{lattice,lattice_mu}
within errors 
the two transitions always appear to coincide, at
least for $T \neq 0$ and small $\mu$.

At large $N_c$, we find a very different phase diagram in the 
$T$-$\mu$ plane, in which the deconfining and chiral transitions split
from one another.
For $\mu \sim 1$, the deconfining transition temperature
is independent of $\mu$.
At low temperatures and densities, there is the usual
confined phase of hadrons, with chiral symmetry breaking.
The confined phase is baryon free, as a Fermi sea of baryons
first forms at a value near the lightest baryon mass.
Within a narrow window in $\mu$, $\sim 1/N_c^2$, there is then
a rapid transition to a dense phase,
with a pressure $\sim N_c$.
The properties of dense phase(s) are illustrated by a Skyrme crystal
\cite{skyrme,skyrme_crystal}.
Although the total pressure is $\sim N_c$, like that of quarks,
the dense phases are confined, 
with interactions near the Fermi surface dominated by baryons.
We suggest that the dense phase undergoes a chiral transition for
$\mu \sim 1$.  In the chirally symmetric phase, the baryons
are parity doubled \cite{parity}, 
consistent with the constraint of anomalies at nonzero density
\cite{anomaly}.

Admittedly, all of our arguments are merely qualitative.
Even so, we think it worth pursuing them, because they are so 
different from naive expectation.
Of course our analysis could simply
be an artifact of the large $N_c$ expansion, and of limited relevance to QCD,
where $N_c = 3$.  At the end, we suggest what our analysis 
might imply about the phase diagram of QCD.

\section{Review of Large $N_c$}

If $g$ is the gauge coupling, the 't Hooft limit is to take
$N_c \rightarrow \infty$, holding $g^2 N_c$ fixed \cite{largeN,witten}.
This selects all planar diagrams of gluons.  Holding
$N_f$ fixed as $N_c \rightarrow \infty$, quark loops are suppressed,
and the only states which survive 
have a definite number of quarks and anti-quarks.

Mesons are composed of one quark anti-quark pair, and are free at infinite
$N_c$: cubic interactions vanish $\sim 1/\sqrt{N_c}$, quartic interactions,
$\sim 1/N_c$, {\it etc.}  Glueballs are pure glue states, with no quarks
or anti-quarks; their cubic interaction vanish $\sim 1/N_c$, and so on.
Except for Goldstone bosons, the lightest bosons have masses 
$\sim \Lambda_{QCD}$.

In contrast, baryons are rather nontrivial.  To form a color singlet,
they have $N_c$ quarks.   Assuming that each quark has an energy
of order $\Lambda_{QCD}$, the mass of a baryon 
$M_B \sim N_c \, \Lambda_{QCD}$ \cite{witten,baryons}.
For instance, a gluon exchanged between any two quarks contributes
$\sim g^2 N_c^2 \sim N_c$, and contributes to an average
Hartree potential.  
Care must be taken with diagrams to higher order:
one includes diagrams which modify the average potential,
but not iterations thereof.
Thus two gluons
exchanged between two different pairs of quarks is $\sim N_c^2$,
but represents an iteration of the average potential.
A diagram which modifies the average potential is given by three quarks
which emit three gluons, and which then interact through a three gluon
coupling, fig. (41) of \cite{witten};
this is $\sim g^4 N_c^3 \sim N_c$.  

The scattering of two baryons begins with
the exchange of two quarks between the baryons, with a gluon
emitted between them,
fig. (35) of \cite{witten}.  Since each quark can have a different
color, this diagram is $\sim g^2 N_c^2 \sim N_c$.
It is also possible to exchange quarks of the same color without
gluon emission; this is also $\sim N_c$, fig. (36) of \cite{witten}.
As for the average potential in one baryon, 
the two body scattering amplitude remains $\sim N_c$ to higher order
in $g^2$.  To see this, it is necessary to pick out contributions
which are two baryon irreducible, from those which 
represent iterations of the two baryon potential.

Continuing, the scattering between three baryons is $\sim N_c$:
if a quark in each baryon emits a single gluon, which interact through
a three gluon interaction, the amplitude is
$\sim g^4 N_c^3 \sim N_c$.  
Again, multiple gluon exchange does not produce higher powers of $N_c$,
once one picks out interactions which are three baryon irreducible.

In general, the scattering of $\cal M$ baryons is of order $\sim N_c$.  
Thus unlike mesons or glueballs, whose interactions vanish at infinite
$N_c$, baryons interact strongly, with couplings of strength $\sim N_c$.

Of course baryon interactions can also be viewed 
as arising from the exchange of color singlet mesons.
The coupling 
between a meson and a baryon $\sim \sqrt{N_c}$, and so 
single meson exchange gives a two baryon interaction
$\sim N_c$, as above.  While it appears that multiple meson exchange
will lead to higher powers of $N_c$, this does not occur,
due to an an extended $SU(2 N_f)$ symmetry \cite{baryons}.  These
cancellations are subtle, and surely have analogies in
nuclear matter.  For our purposes, however, all we require is that
the scattering of $\cal M$ baryons is always of order $\sim N_c$.

The $SU(2N_f)$ symmetry implies
the low energy spectrum of baryons is highly degenerate.  The lowest
mass baryons form multiplets of isospin, $I$, and
spin, $J$ \cite{baryons}.
These multiplets have $I=J$, from $1/2$ to $N_c/2$ for odd $N_c$.
(For QCD, there is one such state, the $\Delta$.)
The splitting in energy between the states in
these multiplets is of order   
$M_B \sim M(1 + \kappa J^2/N_c^2)$, where
$\kappa$ is a constant.  These are the lightest states: there
are other excited baryons with masses $\sim \Lambda_{QCD}$ above
the lightest.

At zero temperature, 
there is no Fermi sea until the chemical potential exceeds the mass of
the fermion.  
Let $M$ be the mass of the lightest baryon, $M \sim N_c$ \cite{witten}.
It is natural to define a ``constituent'' quark mass, 
$$
m_q = M/N_c \; ,
$$ 
which is of order one at large $N_c$.  Thus at $T = 0$, there is no Fermi
sea until the baryon chemical potential
$\mu_B > M$; for the quark chemical potential, this is $\mu > m_q$.

To illustrate how quarks enter at large $N_c$,
consider the gluon self energy at nonzero $T$ and $\mu$.
To lowest order in $g^2$, at zero momentum this is gauge
independent, equal to the square of the Debye mass.  
For $N_f$ massless flavors, its trace equals
\begin{equation}
\Pi^{\mu \mu}(0) = g^2 \left( 
\left( N_c + \frac{N_f}{2} \right) \frac{T^2}{3}
+ \frac{N_f \mu^2}{2 \pi^2}\right) \; ,
\label{debye}
\end{equation}
Taking $N_c \rightarrow \infty$, holding $g^2 N_c$ fixed,
we see that the gluon contribution,
$\sim g^2 N_c T^2 \sim T^2$, survives.  This is the first in an infinite
series of planar, gluon diagrams at infinite $N_c$.
In contrast, whether for $T \neq 0$
and $\mu \neq 0$,
the quark contribution is only $\sim g^2$, and so 
suppressed by $\sim 1/N_c$.  

This is true order by order in perturbation theory, both in vacuum and for
all $T$ and $\mu \sim 1$:
holding $N_f$ fixed as $N_c \rightarrow \infty$, the effects of quarks
loops are suppressed by $\sim 1/N_c$ \cite{largeN,witten}.
This is simply
because there are $\sim N_c^2$ gluons in the adjoint representation,
but only $\sim N_c$ quarks in the fundamental representation.
Since the quark contribution, relative to that of gluons, is $\sim N_f/N_c$, 
it is essential to hold $N_f$ fixed as $N_c \rightarrow \infty$;
{\it i.e.}, to take of limit of large $N_c$, but small $N_f$.

\begin{figure}[ht]

\begin{center} \includegraphics[width=0.60\textwidth]{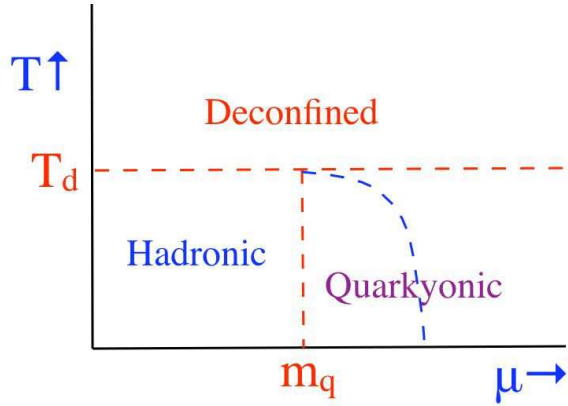}
\caption{Phase diagram at infinite $N_c$ in the plane of temperature
and quark chemical potential.  The blue line in
the quarkyonic phase indicates a guess for
the position of the chiral phase transition.}
\label{phase_diagram}
\end{center} 
\end{figure}

In this limit, we can immediately
make some broad conclusions about
the phase diagram in the $T-\mu$ plane.
At $\mu = 0$, one expects that the deconfining transition temperature
$T_d \sim \Lambda_{QCD}$ \cite{thorn}, which appears to 
be confirmed by numerical simulations on the lattice
\cite{teper}.  Since quarks don't affect the gluons,
the deconfining transition temperature is then {\it independent} of $\mu$,
$T_d(\mu) = T_d(0)$ for values of $\mu \sim 1$.  This
is illustrated in fig. (\ref{phase_diagram}): in the plane of $T$ and $\mu$, 
the phase boundary for deconfinement is a straight line.
The theory is in a deconfined phase when $T>T_d$, and in a confined phase
for $T < T_d$.  

In fact, 
consider the ``box'' in the lower, left hand corner of the
$T-\mu$ plane, where $T < T_d$ and $\mu < m_q$, fig. (\ref{phase_diagram}).
For particles of finite mass,
as long as $T \neq 0$, one expects some population of fermions
in the thermal ensemble.  At large $N_c$, however, the baryons all have
a mass $\sim N_c$; thus if $\mu < m_q$, as long as $T < T_d$, the
relative abundance of baryons is
$\sim \exp(- \kappa N_c)$, 
with $\kappa$ a number of order one,
and so the baryon abundance is exponentially small at large $N_c$.  
(There is a small window in which baryons can be excited, when
$\mu - m_q \sim 1/N_c$, where it costs $\sim 1$, and not $\sim N_c$,
to excite baryons.)

This box, $T < T_d$ and $\mu < m_q$, is the usual, confined phase of
hadrons. At nonzero temperature, the pressure $\sim 1$ is due exclusively
to mesons and glueballs, with only exponentially small contributions
from baryons.  
Even for $\mu > m_q$, the only baryons are those in the Fermi sea,
or excitations thereof.  Simply because they are too heavy,
virtual baryon anti-baryon pairs never contribute at large $N_c$.
This is not true if $\mu$ grows like a power of $N_c$, but
we generally do not consider this regime, except
following eq. (\ref{pressure_Tmu}).

Henceforth we concentrate on cool, dense quarks: 
remaining in the confined phase, $T < T_d$, and moving out in
$\mu$ from $m_q$.  This is the ``quarkyonic'' phase in 
fig. (\ref{phase_diagram}).  We explain this
terminology later, but stress that as it occurs for $T < T_d$
(and $\mu \sim 1$), that it {\it is} confined.
We note that Cleymans and Redlich showed that in QCD,
phenomenologically the boundary
for chemical equilibriation begins at $\mu \approx m_q = M_N/3
= 313$~MeV when $T = 0$ \cite{cleymans}, reminiscent of fig. 
(\ref{phase_diagram}).

\section{Narrow Window of Dilute Baryons}
\label{dilute_baryons}

We start by working at zero temperature, very close to the point
where a Fermi sea of baryons forms.
The Fermi momentum for baryons, 
$k_F$, is $k_F^2 + M^2 = \mu_B^2 $.  If $k_F$ is (arbitrarily)
small, we have an ideal gas of baryons.  For such a gas, the baryon
density is $n(k_F) \sim k_F^3$, the energy is 
$\epsilon \sim k_F^2/2M$, and the pressure is 
\begin{equation}
P_{\rm ideal \; baryons} \sim n(k_F) \; \frac{k_F^2}{M}
\sim \frac{1}{N_c} \; \frac{k_F^5}{\Lambda_{QCD}} \; .
\label{ideal}
\end{equation}
For such a dilute
gas of baryons, the pressure at $T=0$ is very small, $\sim 1/N_c$.

Now consider increasing $k_F$ further, so that the additional resonances
of the baryon condense.  
There are several effects which enter.
The first is to include the resonances with $I=J$, representing the
generalization of the $\Delta$, {\it etc.}, at large $N_c$.
Each species contributes to the pressure as 
$(k_F^2 -(\kappa_I I^2  + \kappa_J J^2)\Lambda_ {QCD}^2)^{5/2}/ M$.
Summing over spin and isospin gives a total contribution of order 
\begin{equation} 
\delta P_{\rm resonances} \sim \frac{1}{M}
\frac{k_F^8}{\Lambda_{QCD}^3} 
\sim \frac{1}{N_c} \; \frac{k_F^8}{\Lambda_{QCD}^4} \; .
\label{resonances}
\end{equation} 
Thus while the sum over resonances changes the dependence upon $k_F$,
it still contributes to the pressure $\sim 1/N_c$.

In contrast, once the nucleon Fermi momentum increases, the effect
of interactions quickly dominates.  The amplitude for
the four point interaction between baryons includes a term
$\sim N_c (\psi^\dagger \psi)^2/\Lambda_{QCD}^2$.
The coupling for this interaction has dimensions of inverse mass squared,
which we assume is typical of $\Lambda_{QCD}$.  At low densities
this interaction contributes of order density squared, or
\begin{equation} 
\delta P_{\rm two \; body \; int.'s} 
\sim N_c \; \frac{n(k_F)^2}{\Lambda_{QCD}^2}
\sim N_c \; \frac{k_F^6}{\Lambda_{QCD}^2} \; .
\label{int}
\end{equation} 
Likewise, six point nucleon interactions contribute as density cubed,
or $\sim N_c k_F^9/\Lambda_{QCD}^5$, {\it etc.}

Now clearly one cannot trust this series when the nucleon Fermi momentum
is of order $\Lambda_{QCD}$.  What is interesting is when the series
breaks down.  Consider balancing (\ref{ideal}) and (\ref{int}): the
two terms are comparable when $k_F^5/N_c \sim N_c k_F^6$, or
\begin{equation}
k_F \sim \frac{1}{N_c^2} \; \Lambda_{QCD} \; .
\end{equation}
Thus at {\it very} low nucleon Fermi momentum, 
$k_F \sim 1/N_c^2$, two body nucleon interactions
are as important as the kinetic terms.  
Contributions from resonances, (\ref{resonances}), are
suppressed by one factor of the 
density, $k_F^3 \sim 1/N_c^6$, as are
three body interactions.

As $k_F$ increases beyond this point,
contributions from the Fermi sea, even including the increasing
number of states, are irrelevant.  Instead, 
when $k_F \sim \Lambda_{QCD}$, the pressure is completely dominated
by baryon {\it interactions}.  Since the interactions between $\cal M$
baryons are $\sim N_c$, all baryon interactions contribute
equally, to give a dense phase in which the pressure is $\sim N_c$.

It is important to stress that at large $N_c$,
the window in which baryons are dilute
is {\it very} narrow.  We expect to enter a dense phase, with
pressure $\sim N_c$, when
the baryon Fermi momentum $k_F \sim 1$.  In terms of the quark
chemical potential, though,
\begin{equation}
\mu - m_q = \frac{\mu_B -M}{N_c} = \frac{k_F^2}{2 M N_c} 
\sim \frac{1}{N_c^2} \;  k_F^2 \; .
\end{equation}  
That is, for $\mu$ one enters a dense regime within
$1/N_c^2$ of the mass threshold.  
This is why in fig. (\ref{phase_diagram}), we have
indicated that the quarkyonic phase begins right at $m_q$: the window
in which one has dilute baryons is only $\sim 1/N_c^2$ in width.

This discussion is somewhat naive.  
If the potential between two nucleons is attractive at large $N_c$ ---
as it is in QCD --- then at arbitrarily low densities free nucleons collapse
to form bound nuclear matter.  
(This is analogous of the tendency of nuclear matter in QCD to go to the
most stable state, which is one of iron nuclei.)
The mass threshold for baryons is then not
at $\mu_B = M_B$, but at $\mu_B = M_B - \delta E$, where 
$\delta E$
is the binding energy of nuclear matter at large $N_c$ \cite{son}.  
One expects $\delta E = N_c \, \delta e$, with $\delta e \sim \Lambda_{QCD}$.
Then in fig. (\ref{phase_diagram}), the quarkyonic phase 
begins not at $m_q$, but within $\sim 1/N_c^2$ of $\mu = m_q - \delta e$.

In QCD, the nuclear binding energy is anomalously small,
$\delta e = \delta E/3 \sim 5$~MeV, versus
$m_q = 313$~MeV.  We do not know
if $\delta e$ is generically
small in the limit of large $N_c$.  If so, it is surely
related to the $SU(2 N_f)$ symmetry \cite{baryons}.  In QCD,
however, this may be an ``accident'' of $2+1$ light flavors.

In QCD, a gas of nucleons doesn't directly collapse to nuclear matter, but
instead exhibits a liquid-gas transition.  This is because of two effects,
the kinetic energy of individual nucleons and Coulomb repulsion.
Neither is important at large $N_c$.  As seen above, 
effects of kinetic energy are automatically $\sim 1/N_c$.
Likewise, taking the baryon electric charge to be of order one at large
$N_c$ (as is necessary for electromagnetism to remain weakly coupled),
then Coulomb repulsion is $\sim 1$, 
and negligible relative to the nuclear potential,
$\sim N_c$.

\section{Dense Baryons as a Skyrme Crystal}
\label{skyrmecrystal}

In the previous section we saw that when the nucleon Fermi momentum
$k_F \sim \Lambda_{QCD}$, that one goes into a dense phase, dominated
by baryon-baryon interactions.  
To understand what might happen in this regime,
in this section we review the properties of Skyrme crystals
\cite{skyrme,skyrme_crystal}.
Our principal interest is as an example of a confined theory
which nevertheless has a chirally symmetric phase.

The usual Skyrme model is a sum of two terms,
\begin{equation}
L = f_\pi^2 \; {\rm tr}
| V_\mu |^2 + 
\kappa \; {\rm tr} [ V_\mu , V_\nu ]^2 \; \;\; , \;\;\;
V_\mu = U^\dagger \partial_\mu U \; \;\; , \;\;\;
U = \exp(i \pi/f_\pi) \; ,
\label{skyrme_action}
\end{equation}
where $f_\pi$ is the pion decay constant, and $\kappa$ a coupling constant,
with $f_\pi^2 \sim \kappa \sim N_c$.  We limit ourselves here to the
Skyrme model for two flavors, where $\pi$ is the pion field.
There are many other terms besides those in (\ref{skyrme_action});
terms from the anomaly contribute
through the Wess-Zumino-Witten (WZW) Lagrangian.  
The two terms above must be viewed as the leading
terms in a derivative expansion.  Terms with higher numbers of
derivatives have coupling constants with
dimensions of inverse mass squared.
The mass dimension of these other terms is presumably set by
(inverse) powers of $\Lambda_{QCD}$.

A single Skyrmion is given by a solution to the field equations from
(\ref{skyrme_action}) over all of space-time.  Since the terms in the
action are $\sim N_c$, the energy of a configuration is also of the same
order, and represents a single baryon.

As shown by Klebanov \cite{skyrme_crystal}, 
a realistic crystal is given by considering periodic solutions in
a finite box.  Like the energy of a single baryon, the energy
of the Skymrion crystal is automatically $\sim N_c$, with one
baryon per box. 
Solving the Skyrme equations of motion for a system with cubic
symmetry is technically involved.
For many crystals, however, it is known that a reasonable approximation is
to chop off the corners of the cube, and to consider the theory on a 
sphere.  This approximation was adopted by Kutschera, Pethick,
and Ravenhall, and also by Manton \cite{skyrme_crystal}.  Many
properties of the crystal with cubic symmetry
are especially transparent for a spherical geometry.  

If $R$ is the size of the sphere, the solution is constructed so that
there is one baryon per sphere, so the baryon density is
$1/(4 \pi R^3/3)$.  A crystal is a bad approximation for large $R$,
but presumably reasonable when
$R\sim \sqrt{\kappa}/f_\pi$.  At large $N_c$, this mass scale
$\sim 1$.

For large spheres, the chiral symmetry is broken, as the $U$ field points
in a given direction in isospin space; typically, $\pi \rightarrow 0$
at spatial infinity.  As $R$ decreases, the stationary point is distorted
by the finite volume of the sphere.

As the radius of the sphere becomes small, there is a phase transition
to a chirally symmetric phase.  For small spheres, the stationary point
is just the identity map, from
from $S^3$ of the $U$'s to $S^3$ of space, taking $\pi^a \sim \hat{r}^a$.
This has unit baryon number per spherical volume, but it is also easy to
see that the integral of $U$, over the sphere, vanishes.

The restoration of chiral symmetry is less obvious for a crystal
with cubic symmetry.  It follows from the half-Skyrmion symmetry
of Goldhaber and Manton, where
the total chiral order parameter cancels
between different regions of the crystal \cite{skyrme_crystal}.  
(For more than two flavors, the correct representation of the chirally
symmetric phase is presumably a generalization 
of this half-Skyrmion symmetry.)

The crucial test for the restoration of chiral symmetry is that the excitation
modes fall into chiral multiplets of the unbroken symmetry.
This was shown for the mesonic excitations of the crystal by
Forkel {\it et al.} \cite{skyrme_crystal}. 
Since the baryon current is topological, the
baryon excitations are not evident.  
In a chirally symmetric phase, they must be parity doubled.
In detail, this happens because the baryon is a topological current.
Thus it is given by integrating over the entire box; if the configuration
is chirally symmetric, so are the integrals thereof.

The Skyrme crystal does not give one insight into all properties of the
system.  The pressure of the system is not obvious, nor even the 
chemical potential of baryons.  What one can do is to compute
how the energy, per cell, depends upon the density.
{\it If} one takes only the two
terms in the Skyrme Lagrangian of (\ref{skyrme_action}), then the term
with four derivatives dominates at small $R$, which is high density.  Since
this term is scale invariant, one automatically finds that the 
relationship between the energy density, $e$, and the density, $n$,
is that for a conformally symmetric theory, $e \sim n^{4/3}$, controlled
by the coupling $\kappa$.

This is an accident of keeping only two terms in the Skyrme lagrangian.
Terms with six derivatives, for example, are also proportional to $N_c$,
with dimensions of $1/\Lambda_{QCD}^2$.  When the size of the crystal
is $\sim 1/\Lambda_{QCD}$, however, {\it all} such interactions are equally
important.  This is analogous to the counting for baryon baryon interactions,
which are always $\sim N_c$, and which are characterized by mass scales
$\sim \Lambda_{QCD}$.

We stress that the Skyrme crystal is only meant 
as an illustration.  For example,
while naively one expects that a system
of heavy particles forms a crystal, since the interactions are as large,
also $\sim N_c$, even this may not necessarily follow.  

\section{Quarkyonic Matter at Large $N_c$}
\label{cool_dense}

Skyrme crystals illustrate what might happen as $\mu$ increases from
$m_q$ (or $\mu_B$ from $M$).  
Alternately, one can consider working down in $\mu$.
For very large $\mu$, $\mu \gg \Lambda_{QCD}$, one should be able
to compute the total pressure by perturbation theory, in the QCD
coupling $g^2$:
\begin{equation}
P_{\rm pert.}(\mu) \sim N_c N_f \;\mu^4 
\; F_0(g^2(\mu/\Lambda_{QCD}),N_f) \; ,
\label{perturbative_pressure}
\end{equation}
The function $F_0$ 
has been computed to $\sim g^4$ \cite{dense}.  
At large $N_c$, it is given by planar diagrams with a single quark
line, and an arbitrary number of gluon insertions.
The pressure of ideal (massless) quarks is $\sim \mu^4$.
Perturbative corrections are given by a gluon
knocking a quark in the Fermi sea out of it, the 
subsequent rescattering of the quark(s) and gluons, {\it etc.}
When $\mu \gg \Lambda_{QCD}$, for quarks deep in the Fermi sea, 
all such contributions are due to hard scattering, with the energy and
momenta transferred $\sim \mu$.
Since the gluon has momentum components
of order $\mu$, the the coupling runs according to this scale,
with the $\beta$-function of the pure glue theory,
(\ref{perturbative_pressure}). 
Thus in perturbation theory,
the pressure is a power series 
in $\sim 1/\log(\mu/\Lambda_{QCD})$, {\it etc.} times $\mu^4$.

To be able to compute the total pressure reliably in perturbation theory,
we only need to assume that $\mu \gg \Lambda_{QCD}$.  In particular,
it is {\it not} necessary to assume that $\mu$ grows like a power of
$N_c$.  Presumably a perturbative calculation is applicable for,
{\it e.g.}, $\mu > 10^2
\Lambda_{QCD}$.  When $N_c$ is absurdly large,
such as $N_c = 10^{12}$, this is still smaller than any other scale which
enters at large $N_c$, such as 
$N_c^{1/4} \Lambda_{QCD} \sim 10^3 \Lambda_{QCD}$; see the discussion
following eq. (\ref{pressure_Tmu}).  

Now consider pushing the perturbative computation of the pressure
down to $\mu \sim \Lambda_{QCD}$.  
Non-perturbative contributions to the pressure enter, through terms
such as $\sim \mu^2$, eq. (\ref{fuzzybag}).  Even so, outside of the
window of dilute baryons, sec. \ref{dilute_baryons}, baryons are
dense, and we expect that the pressure remains $\sim N_c$.

This then raises
the central conundrum of our work: 
for $T < T_d$, and $\mu \sim 1$, it appears that we can describe
the system {\it either} as one of confined baryons, {\it or} as one of quarks.
Admittedly, the baryons interact strongly, $\sim N_c$; likewise,
and especially 
for $\mu \sim \Lambda_{QCD}$, the Fermi sea of quarks is far from ideal.  
But how can {\it both} pictures apply?

We suggest the following resolution.  At large $\mu \gg \Lambda_{QCD}$, 
for quarks
far from the Fermi surface, their scattering can be reliably computed
in perturbation theory.  This is reasonable: at large $\mu$,
the density of quarks per hadronic volume
$\sim 1/\Lambda_{QCD}^3$, is large.  In
such a dense medium, a quark
doesn't know which baryon it belongs to; then, for
the most part, it is appropriate to
view the system as one of (non-ideal) quarks.

Even at large $\mu$, however, 
it is essential to consider separately the scattering of particles
within $\sim \Lambda_{QCD}$ of the Fermi surface.
In this regime, quarks interact by exchanging 
gluons with momenta $\sim \Lambda_{QCD}$.  At infinite $N_c$, where quarks
cannot screen gluons, we know how quarks at $\mu \neq 0$ scatter: 
exactly as for $\mu = 0$.  When $T < T_d$, then,
the theory is in a confined phase, and so near the Fermi surface, 
it is appropriate to speak not of the scattering of quarks, but of baryons.

We term this a 
{\it ``quarkyonic''} phase: a quark Fermi sea, with a baryonic
Fermi surface.  The width of the baryonic surface is $\sim \Lambda_{QCD}$,
so when $\mu \sim \Lambda_{QCD}$, it is all baryons. 
As $\mu$ increases, the baryons form a band, of approximately
constant width, on the edge of the
Fermi surface.  
There is no
quantitative difference between a quarkyonic
phase, with a wide baryon surface, and one with a narrow surface:
they smoothly interpolate from one to the other.
At large $N_c$, it is possible to differentiate the quarkyonic phase,
with a pressure $\sim N_c$, from that in the hadronic phase, $\sim 1$,
or the deconfined phase, $\sim N_c^2$.
This clear distinction is only possible at large $N_c$ (and small $N_f$).

The effects of a baryonic Fermi surface
show up in the pressure through terms which are powers of 
$\sim (\Lambda_{QCD}/\mu)^2$ times the ideal gas term,
eq. (\ref{fuzzybag}).
This is typical of a nonperturbative correction, as an inverse power
of a (hard) mass scale.  When $\mu \sim \Lambda_{QCD}$, this
is a large correction.  
When $\mu \gg \Lambda_{QCD}$, numerically this is a very small
contribution to the total pressure.  Even so, as particles at
the edge of the Fermi surface are the lightest excitations, 
even at large $\mu$ baryons dominate processes with low momenta.
This implies that at large $N_c$,
phenomena involving the Fermi surface, such as
superconductivity and superfluidity, are properly described by
baryons, and not by quarks, for {\it all} $\mu \sim 1$.

By considering gluonic probes, it is 
clear that the theory is in a confined phase for $T < T_d$ and $\mu \sim 1$.
As discussed by Greensite and Halpern \cite{thorn}, at zero temperature
the Wilson loop is insensitive to quarks at large $N_c$:
it exhibits an area law, with a nonzero string tension.
Screening due to quarks enters through corrections $\sim 1/N_c$.
Adding a Fermi sea of quarks doesn't change this,
as long as $\mu \sim 1$.

At nonzero temperature, the order parameter for deconfinement
is the renormalized Polyakov loop \cite{banks_ukawa,bielefeld_ren_loop}.  
Quarks induce an expectation value for the Polyakov loop,
but this is $\sim 1/N_c$, versus a value $\sim 1$ in the deconfined phase.
Thus up to corrections $\sim 1/N_c$, as a function
of temperature the expectation value
of the renormalized Polyakov loop
is independent of $\mu$ (for $\mu \sim 1$);  {\it e.g.},
in fig. (\ref{phase_diagram}), $T_d(\mu)$ is a straight line.

That the theory confines can also be seen by exciting a quark,
in the Fermi sea, with some external probe.
If the quark is deep in the Fermi sea, knocking it out takes
a probe with large momentum.  When $\mu \gg \Lambda_{QCD}$, at first
the resulting quark propagates like a hard quark.
It, and the remaining hole in the Fermi sea, then scatter off of other
quarks in the Fermi sea, knocking some out, creating other holes, and so
on.  Eventually, one ends up with not a single, unconfined quark,
but a perturbed Fermi sea, characterized by some number of
excited {\it baryons} and their holes.  That the theory confines is
clearer if the external probe carries momentum $\sim \Lambda_{QCD}$:
then one immediately sees that the only particles (and holes)
excited near the Fermi surafce are not quarks, but baryons.

The above discussion applies to quarks of any mass, and
leads to the phase diagram of fig. (\ref{phase_diagram}).  An example
is provided by the solution of QCD in
$1+1$ dimensions \cite{largeN}.  In two dimensions
there is only a confined phase, $T_d = \infty$.
Schon and Thies showed that at nonzero quark density,
the quark propagator remains infrared divergent, and so confined,
as it is in vacuum \cite{schon_thies}.
This is exactly what one expects of a quarkyonic phase: that only
color singlet excitations, such as mesons and baryons, have finite
energy.  
It would be interesting to perform more detailed calculations,
such as of the free energy, and how the properties of
mesons and baryons change with $\mu$.

Returning to four dimensions, the crucial question is: 
where is the chiral phase transition for light quarks?
For $\mu < m_q$, it surely coincides with the deconfining phase
transition, and occurs at $T_d$.  We suggest, however, that the
two transitions no longer coincide when $\mu > m_q$.
If we take the Skyrme crystal as a guide to the quarkyonic phase,
then chiral symmetry restoration occurs
not at the mass threshold, but above $\mu = m_q$.  This is because
there must be some significant density of baryons in the Fermi sea to drive
the transition.  Further, the transition
occurs when $\mu \sim m_q$, on the order of the constituent quark mass,
and not at some $\mu$ which is asymptotically
large in a power of $N_c$.
Schematically, this gives the blue line in fig. (\ref{phase_diagram}).

This can also be seen in illustrative models.
As an example of a possible
solution to the Schwinger-Dyson equations \cite{sch_dyson},
Wagenbrunn and Glozman \cite{parity}
studied a model with a confining gluon propagator, $\sim 1/(k^2)^2$ in
momentum space.  At infinite $N_c$, solutions
to the Schwinger-Dyson equations are those
where the quark propagator (and its vertex with gluons)
change, but the gluon propagator doesn't.  Computing with the quark
propagator at $\mu \neq 0$, but leaving the gluon propagator 
unchanged, it is not difficult to 
see that increasing $\mu$ drives chiral symmetry restoration 
at some nonzero value of $\mu$ above the mass threshold.

This can also be seen from the eigenvalue distribution of the Dirac operator.
In vacuum, Banks and Casher showed that
chiral symmetry breaking is driven by a nonzero density
eigenvalues at zero eigenvalue
\cite{anomaly}.  When $\mu \neq 0$, for
$N_f \geq 3$ the eigenvalues spread out in the complex plane, and there
is no simple analogous condition \cite{random}.  
Even so, it is most natural then
as $\mu$ increases above the mass threshold,
that whatever effect the gauge fields have on the eigenvalues, that
eventually it is overwhelmed by $\mu \neq 0$.
An explicit example of this
is chiral symmetry restoration in random matrix models \cite{random}.

The splitting of the deconfining and chiral transitions can be 
represented naturally in effective models.  Mocsy, Sannino, and
Tuominen \cite{two_trans} showed that if the coupling between the
Polyakov loop and the chiral condensate has one sign, the transitions
coincide; for the other, they diverge.  Thus this coupling vanishes
at the point where the transitions diverge.

At large $N_c$, because of the changes in the magnitude of the pressure,
the deconfining transition is expected to be of first order.  Once
the chiral transition no longer coincides with deconfinement,
its order is presumably controlled by the usual renormalization group
analysis \cite{two_trans}, and depends strongly upon the number of
flavors.  

There is one caveat which must be noted.  Chiral symmetry,
and its possible restoration, can be sensitive to pairing near the
Fermi surface.  Certainly for $\mu \gg \Lambda_{QCD}$, one expects
that quarks deep in the Fermi sea are best described as chirally
symmetric.  It is possible, however, that the baryons near the Fermi
surface may experience non-perturbative effects which cause them to
pair in a chirally asymmetric manner.  This effect will manifestly be
small, suppressed at least by $\sim (\Lambda_{QCD}/\mu)^2$.

Up to this point, 
we have assumed that $\mu \sim 1$, so that gluons
are blind to quarks and their Fermi sea.  Especially to understand 
finite $N_c$, however, it is also necessary to
consider values of
$\mu$ which grow with (fractional) powers of $N_c$.  
We assume that the temperature $T \sim \Lambda_{QCD}$, like $T_d(0)$,
the temperature for deconfinement at $\mu = 0$.
In perturbation theory, the pressure includes terms as
\begin{equation}
P_{\rm pert.}(\mu,T) \sim N_c N_f \;\mu^4 \; F_0 \;\;\; , \;\;\;
N_c N_f \; \mu^2 \; T^2 \; F_1 \;\;\; , \;\;\;
N_c^2 \; T^4 \; F_2 \; .
\label{pressure_Tmu}
\end{equation}
In the limit of large $N_c$, 
$F_0$, $F_1$ and $F_2$ are functions of the coupling constant 
$g^2$ and $N_f$.
The coupling runs with both mass scales, $\mu$ and $T$.
Thus when $\mu$ grows like a power of $N_c$ times $\Lambda_{QCD}$,
perturbation theory in $g^2$ is a good approximation.
As always, this is excepting power like
corrections from the region near the Fermi surface.

Consider $\mu \sim N_c^{1/4} \Lambda_{QCD}$.  
In this region, the quark contribution to the
pressure, $\sim N_c \, \mu^4 \, F_0 \sim N_c^2 \, F_0$, is as large as that of
deconfined gluons,
$\sim N_c^2 \, F_2$. In this regime, the quark contribution to the pressure
is independent of temperature, 
since $\sim N_c \, \mu^2 \, F_1 \sim N_c^{3/2} \, F_1$ 
is down by $\sim 1/\sqrt{N_c}$.  
It is possible that there is a temperature dependent term in the pressure,
induced by baryons near the Fermi surface; {\it e.g.},
$\sim N_c \, N_f \, T^2 \, \Lambda_{QCD}^2 $, 
which is down by $\sim 1/N_c$ to the leading term, $\sim N_c^2$.

At larger values of 
$\mu \sim N_c^{1/2} \Lambda_{QCD}$, quarks contribute to the Debye mass
in the limit of large $N_c$, 
eq. (\ref{debye}).  In this regime, the pressure
is completely dominated by that of quarks at zero temperature,
$\sim N_c\, \mu^4 \,F_0 \sim N_c^3 \, F_0$.  In perturbation theory, the gluon
contribution to the pressure remains
$\sim N_c \,T^4 \, F_2 \sim N_c^2$, with
the temperature dependent part of the quark
pressure also $\sim N_c \, \mu^2 \, T^2 \, F_2 \sim N_c^2$;
thus both are down 
by $1/N_c$, relative to the quark term at zero temperature.  

When $\mu \sim N_c^{1/2} \Lambda_{QCD}$, 
gluons are screened by quarks, and one is
in a qualitatively new regime.  At zero temperature, the Wilson
loop no longer exhibits an area law;
at nonzero temperature, the renormalized Polyakov loop
acquires an expectation value of order one.  Consequently, eventually
the first order phase transition for deconfinement ends.  It
can do so in one of two ways: the phase
boundary for deconfinement can either bend over to zero
temperature, or it can end in a critical end point at $T \neq 0$.
Our simple arguments cannot predict which occurs.
It is reasonable that either is only possible once gluons feel
the quarks; {\it i.e.}, when
$\mu \sim N_c^{1/2} \Lambda_{QCD}$.  
If so, and there is a critical end point,
the critical behavior is a correction
in $1/N_c$ to the total pressure, which is dominated by
the zero temperature term for quarks.

We remark that if $N_f$ goes to infinity with $N_c$, then all
terms in eq. (\ref{pressure_Tmu}) are $\sim N_c^2$.
Indeed, when both $N_c$ and $N_f$ are large, then even in vacuum
one cannot speak, rigorously, of confinement.
We have no insight into this limit.

Deryagin, Grigoriev, and Rubakov showed that at large $N_c$,
the color singlet pairing of chiral density waves
dominates over the di-quark pairing
of color superconductivity \cite{chir_dens}.
Their perturbative analysis is reliable for
$\mu \gg \Lambda_{QCD}$, as long as the quarks do not lie
within $\sim \Lambda_{QCD}$ of the Fermi surface.  When they do,
the pairing of baryons also contributes.

Son and Shuster \cite{chir_dens} showed that 
even for extremely large values of $N_c$, Debye screening
disfavors the pairing to chiral density waves, and quark color
superconductivity dominates.
In sec. (\ref{quarkqcd}), we adopt a similar criterion to
estimate when in QCD there is a 
transition from a quarkyonic, to a perturbative, regime.

\section{Confinement and Chiral Symmetry Breaking}
\label{conf_chiral}

At zero temperature and density, the effect of anomalies usually ensures
that any confined phase is one in which chiral symmetry is broken.  
We now give a heuristic argument as to
why this need not be true at nonzero density. 

Casher, and then Casher and Banks, argued that in the vacuum, confinement 
automatically implies the breaking of chiral symmetry \cite{anomaly}.
Consider a meson, in which the quark 
propagates to the right, with a spin along its direction of motion.  
To remain a meson at rest, this must mix with a quark propagating to the
left, which can happen by scattering off of a gluon.  
Its spin, however, is now opposite to the direction of motion, so its
helicity has been flipped.  Since in QCD the interactions preserve chirality,
which for a massless field equals helicity, this change of direction cannot
occur.  It can if there is a mass condensate in the vacuum, which the quark can
scatter off of, and flip its helicity.  Note that this
argument is especially tight in the limit of large $N_c$, where the number
of quarks in a meson is fixed.  

Now consider the similar process at nonzero temperature.  Then besides 
scattering off of a gluon, one can scatter of a quark in a thermal 
distribution.  However, if we consider the processes of both emission
and adsorption, the total is
$\widetilde{n}(E_k) - \widetilde{n}(E_{k'})$;
$E_k$ is the energy of the rightgoing quark, is $E_{k'}$ is the
energy of the leftgoing quark, and $\widetilde{n}(E)$ is the Fermi-Dirac
statistical distribution function.

For an isotropic distribution, as in thermal equilibrium, if the momenta
are the same, then the two distribution functions cancel. 
Thus the process is only
allowed when $k \neq k'$.  Since the momenta of the right and left moving
quarks are different, however, we end up with an excited meson, different
from the initial meson.  That is, this process represents not a meson
at rest, but scattering between a meson, and some thermally excited state,
such as another meson.  

In general, this is fine in a thermal distribution: what we mean by a 
``meson'' is a sum over states anyway.  However, there is a problem
at large $N_c$: if $T \neq 0$ and $\mu = 0$, then all interactions vanish
at large $N_c$, and this process must be suppressed by powers of $1/N_c$.
Thus Casher's argument suggests that at nonzero temperature, and $\mu = 0$,
the connection between chiral symmetry and confinement remains.

This connection could be lost in the presence of a Fermi sea, however.
The argument goes through as before, except now we scatter off of a 
quark in the Fermi sea.  Physically, the quark in the test meson scatters
off a quark in a baryon, which then scatters into a baryon hole.
There is no inconsistency with the large $N_c$ expansion, because the
scattering amplitude is large, of order one.  
This suggests that it is possible to have a confined, but
chirally symmetric, phase at $\mu \neq 0$.

What of the constraints from anomalies, which after all, are due to
ultraviolet effects?  Certainly the anomaly itself is unchanged
by temperature or density.  However, their implications are less
obvious when $T$ or $\mu$ are nonzero.  Because of the breaking
of Lorentz invariance at $T$ and $\mu \neq 0$, Itoyama and Mueller
showed that many more amplitudes arise \cite{anomaly}.
The anomaly relates these amplitudes, but not as directly as in vacuum.
For example, Pisarski, Trueman, and Tytgat showed that the
Sutherland-Veltman theorem, which relates 
the amplitudes for $\pi^0 \rightarrow \gamma \gamma$,
does not apply at nonzero temperature \cite{anomaly}.
Thus the connection between chiral symmetry breaking, and confinement,
need not remain at nonzero $T$ or $\mu$.

The above generalization of Casher's argument suggests that a system
at $\mu \neq 0$ is uniquely different from $\mu = 0$ and $T \neq 0$.
This may arise as follows.  
Anomalies are saturated by excitations with arbitrarily low energies;
for example, when chiral symmetry breaking occurs, by pions.
To model a chirally symmetric
phase, consider massive, parity doubled baryons \cite{parity}.
In a thermal distribution with
$\mu = 0$, massive modes are Boltzmann suppressed, and cannot be
excited at low energy.
At nonzero density, however, a Fermi
sea of massive particles can be excited, with arbitrarily small
energy, by forming a particle hole pair.

We conclude this section by noting that the Skyrme model provides
a direct example of a confined theory which satisfies the anomaly
conditions.  There, the
Wess-Zumino-Witten term automatically incorporates all effects of the anomaly,
such as $\pi^0 \rightarrow \gamma \gamma$, {\it etc.}  This happens whether
the background field is chirally asymmetric, as for large $R$, or chirally
symmetric, as for small $R$.  In either case, 
fluctuations from the Wess-Zumino-Witten terms
automatically incorporate all anomalous amplitudes.

\section{Quarkyonic Matter in QCD}
\label{quarkqcd}

The large $N_c$ limit we consider is only simple because $N_f$
is held fixed as $N_c \rightarrow \infty$.
Since $N_c = N_f = 3$ in QCD, then
especially at $\mu \neq 0$, it is far from clear that QCD
really is close to this limit of large $N_c$, and small $N_f$.
For the purposes of discussion, we 
henceforth assume that it is.

To extend the analysis at large $N_c$ to QCD, the principal physical
effect to include is that of Debye screening.
From (\ref{debye}), the Debye mass is
$m_{Debye}^2 = (2 N_f/\pi) \alpha_s(\mu) \mu^2$, where
$\alpha_s = g^2/(4 \pi)$.
This is to be compared with
the scale of confinement, $\Lambda_{QCD}$.  The latter is only approximate:
probably a better measure of the confinement scale is not $\Lambda_{QCD}$
{\it per se}, but the mass of the $\rho$ meson, $\approx 1$~GeV.

The Debye mass can be computed, at zero temperature and nonzero density,
to higher order in perturbation theory \cite{dense}.  
The really essential question is to know how
the effective coupling runs.  At nonzero temperature, and $\mu = 0$,
Braaten and Nieto suggested in the imaginary time formalism, as energies
are always multiples of $2 \pi T$, perhaps the effective coupling
runs in the same way \cite{pert_temp}.  This was confirmed
by computations to two loop order \cite{pert_temp}.
This implies that while $T_d \sim 200$~MeV is relatively low, that 
the effective coupling is moderate in strength, 
even down to $T_d$ \cite{banks_ukawa}.

There doesn't appear to be any similar factor at nonzero density; at
$T = 0$, the coupling should run like 
$\alpha_s(\mu/(c\Lambda_{QCD}))$, where
$c$ is a number of order one.  
For purposes of discussion, we that assume 
perturbation theory is reliable for $\mu > 1$~GeV; at this
scale, the Debye mass is also $\sim 1$~GeV.

A Fermi sea first forms when the quark chemical potential
$\mu > M_N/3 \approx 313$~MeV.  
Large $N_c$
suggests that dilute baryons persist only in a narrow window, 
$\sim \Lambda_{QCD}/N_c^2$.  Then, at some scale above this,
QCD becomes quarkyonic, in that the pressure rises rapidly.
Notice that the increase in pressure is {\it not} associated
with a phase transition.  In terms of baryons, it appears to
be due entirely to their strong interactions.  
Below $\mu \sim 1$~GeV, it can also be viewed as due to the strong
interactions amongst highly non-ideal quarks.

Once Debye screening becomes significant above $\mu \sim 1$~GeV,
gluons are shielded, and the coupling may be (relatively) moderate in
strength.  Because of Debye screening, at large $\mu$ scattering within
$\Lambda_{QCD}$ of the Fermi surface should be under control.  This is
unlike large $N_c$, where Debye screening doesn't contribute until
$\mu \sim \sqrt{N_c}$.  The transition from a quarkyonic regime, to
one which is perturbative in quarks and gluons, is presumably smooth,
as there is no order parameter to distinguish one from the other.
(Assuming that the deconfining transition doesn't extend down to $T = 0$,
see below.)

Needless to say, our estimates for $\mu$ in the quarkyonic phase
are {\it extremely} crude.
Understanding the lower bound on $\mu$ requires
matching onto models of nuclear matter; perhaps it might help by
matching onto models consistent with large $N_c$ counting.
The upper limit can be pinned down better through higher order
calculations in perturbation theory
at $\mu \neq 0$ and $T = 0$ \cite{dense}.

\begin{figure}[ht]
\begin{center} \includegraphics[width=0.60\textwidth]{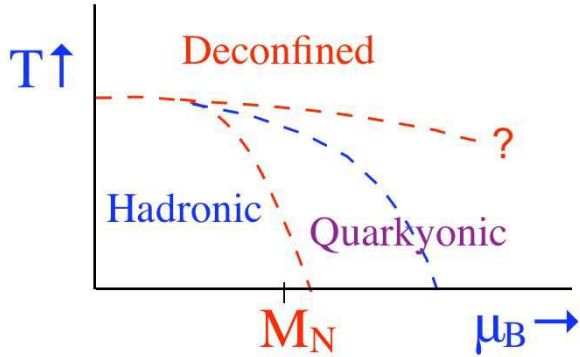}
\caption{Possible phase diagram for QCD in the plane of 
temperature and baryon chemical potential.  The blue line in
the quarkyonic phase indicates the chiral phase transition.
There is a critical end point for deconfinement.}
\label{QCDphase_diagram}
\end{center} 
\end{figure} 

A possible phase diagram is drawn in fig. (\ref{QCDphase_diagram});
following phenomenology \cite{cleymans}, we plot this as
as a function of the temperature and the baryon chemical potential,
$\mu_B$.  
If large $N_c$ is a reasonable guide to $N_c = 3$, this should look
something like fig. (\ref{phase_diagram}), except that the sharp edges
are smoothed out.  For example, below the mass threshold, $T_d$
should change little with $\mu$; this appears to be true from 
numerical simulations on the lattice \cite{lattice_mu}.
Similarly, at large $N_c$ nuclear matter rapidly goes from a dilute phase,
to one which is dense and quarkyonic.  We indicate this in the figure
by drawing the quarkyonic phase slightly above $M_N$, the nucleon mass.  

We expect that the chiral phase transition occurs in the quarkyonic 
phase, well 
above the mass threshold.  For QCD, at present numerical simulations
on the lattice indicate that for small $\mu$, the deconfining 
and chiral transitions coincide, and are crossover.  A chiral critical
end point may exist in the plane of $T$ and $\mu_B$ \cite{tricritical}.
One might conjecture that if such a critical end point exists, that the
deconfining and chiral transitions split from one another at that point.

Speculating in this manner, in the quarkyonic phase,
the latent heat associated with the chiral
transition might be relatively small.  Certainly at large $N_c$, the
large increase in pressure, $\sim N_c$, is not tied to the chiral transition.
The behavior of the chiral transition is very sensitive to the number
of flavors, and possible restoration of the axial $U(1)$ symmetry, though.

Consider the deconfining phase transition, after it splits from the
chiral transition.  At fixed $\mu$, as $T$ increases, one goes from a
confined phase of parity doubled baryons, to one of quarks and gluons.
Deconfinement could either remain crossover, or perhaps become first
order again (from the splitting point?).  
If it does turn first order, it will then have to end
in a critical end point, now for deconfinement.  Alternately, 
a first order deconfining transition could perist down to zero temperature.
We indicate this uncertainty
by the question mark in fig. (\ref{QCDphase_diagram}).

How can the quarkyonic phase be studied?  
For the total pressure, one should use a description not in
terms of baryons, but in terms of quarks.
Admittedly, they are highly non-ideal quarks, but there are hints to
their possible behavior from numerical simulations, on the lattice, at nonzero
temperature.  At $T \neq 0$ and $\mu = 0$, 
the pressure can be characterized by a generalized (or ``fuzzy'') bag
model.  This is a 
power series in $1/T^2$ times the ideal gas term 
\cite{bielefeld96,lattice,banks_ukawa}.
At $T=0$, and nonzero $\mu$, this suggests 
\begin{equation}
P_{\rm quarkyonic}(\mu) 
= f_{\rm pert} \; \mu^4 - \mu_c^2 \; \mu^2 - B + \ldots 
\label{fuzzybag}
\end{equation}
Perturbative corrections are subsumed into $f_{\rm pert}$.  
Nonperturbative corrections, such as due to confinement, are included in
$\mu_c^2$ and $B$.  Because of the term $\sim \mu_c^2 \mu^2$, the constant $B$
need not agree with the usual MIT bag term, even in sign.
We note that such a parametrization arises naturally from a Skyrme crystal.  
In the simplest model, what is equivalent to a conformally symmetric
term $\sim \mu^4$ arises from the Skyrme term, $\sim \kappa$.  Power
like corrections then arise from the usual sigma Lagrangian, 
$\mu_c^2 \sim f_\pi^2$, {\it etc.}  The pressure from this
generalized bag model should then match 
{\it smoothly} onto
that of nuclear matter, with no phase transition between the two.

In contrast, in order to compute properties near the Fermi surface, it is
necessary to consider effective theories of baryons.
In a phase with chiral symmetry breaking, at low density
these must match onto models of nuclear matter.
In a chirally symmetric phase, the baryons are parity doubled.  
One possibility is to use 
Nambu-Jona-Lasino (NJL) models, not of quarks \cite{cs_crystal,njl}, 
but of baryons.  Linear models of parity doubled baryons may also
be of use \cite{parity}.
Phenomenon such as superfluidity and superconductivity, and transport
properties in general, are dominated by these states.
For parity doubled baryons, the patterns of baryonic superfluidity and
superconductivity will be significantly constrained by 
anomaly conditions.  One might guess that the scales of baryon pairing in
the quarkyonic phase is on the order of those
in ordinary nuclear matter; {\it i.e.}, that the gaps are small,
tens of MeV.

Sch\"afer and Wilczek \cite{cs_pert} noted that 
for three light flavors, there is continuity
between a nucleonic phase and one with quark color superconductivity.
While chiral symmetry breaking is large in a nucleonic phase, it is
also generated by color-flavor locking.
This suggests that for quarkyonic matter
with three flavors and three colors, that one
possibility is for the massive, parity doubled baryons to form 
(small) gaps which spontaneously break chiral symmetry.
This is the simplest way
by which massive baryons, which are now only approximately
parity doubled, can satisfy anomaly constraints at $\mu \neq 0$,
although we suspect there are others.

One way of computing the properties of a quarkyonic phase is to use
approximate solutions of Schwinger-Dyson equations \cite{sch_dyson}.
These are, almost uniquely, the
one approximation scheme which includes both confinement and chiral
symmetry breaking.  They do have features reminiscent of large $N_c$:
at low momentum, if chiral symmetry breaking occurs, the gluon
propagator for $N_f = 3$ is numerically close to that for $N_f = 0$.
At present, solutions at $\mu \neq 0$ assume a Fermi surface dominated
by quarks; if quark screening is not too large
at moderate $\mu$, these models should exhibit a quarkyonic phase.

On the lattice, it 
is well known that while gauge theories have a sign
problem at nonzero quark density when $N_c \geq 3$, that 
numerical simulations can be done for two colors.
Recently, these were done 
at $T=0$ and $\mu \neq 0$ for heavy quarks: they
exhibit superfluidity at the mass threshold and deconfinement
well above it \cite{two_colors}.
These simulations could be extended to
light quarks, to see if the phase diagram is anything like that of 
fig. (\ref{QCDphase_diagram}): {\it e.g.}, at low temperature, is
chiral symmetry restored before deconfinement?

While our analysis is crude, existing prejudice has been that the 
phase transitions for deconfinement and chiral symmetry
are inexorably linked together.  Large $N_c$ suggests that by moving
out in chemical potential, that potentially one has the chance to see
the two transitions separate.  This could happen at rather high temperature,
near that for the deconfining transition 
temperature at zero density,
and relatively low density, less than that for nuclear matter.
Experimentally, it is possible to move out in the plane in $\mu$,
at high $T$, by going to ``low'' energies,
such as at critRHIC and FAIR.  Thus these facilities
may explore not just a chiral critical end point
\cite{tricritical}, but quarkyonic phases, including one
which is confined, yet chirally symmetric.

\section{Acknowledgements} This work was supported by the U. S.
Department of Energy under contract number DE-AC02-98CH10886.
We thank R. Alkofer, T. Cohen, K. Fukushima, L. Glozman, A. D. Jackson, 
A. Mocsy, C. J. Pethick, P. Petreczky, P. Romatschke, 
J. Schaffner-Bielich, K. Schwenzer, D. Son, K. Splittorff,
R. Venugopalan, and J. Verbaarschot for comments and discussions.
R.D.P. would also like to thank 
the Alexander von Humboldt Foundation for their support, 
and Prof. R. Alkofer, at the University of Graz, 
and Dr. P. Damgaard, at the Niels Bohr Institute, for their
generous hospitality.

\end{document}